\documentclass[%
 reprint,
 amsmath,amssymb,
 aps,
]{revtex4-2}

\usepackage{yfonts}
\usepackage{xcolor}
\usepackage{xcite}
\usepackage{braket}
\usepackage{graphicx}
\usepackage{dcolumn}
\usepackage{bm}

\begin{document}

\preprint{APS/123-QED}

\title{
An Extended Second Law of Thermodynamics}

 \author{Alejandro Corichi}
 \email{corichi@matmor.unam.mx}
 \affiliation{%
Centro de Ciencias Matemáticas, Universidad Nacional Autónoma de México, UNAM-Campus Morelia, A. Postal 61-3, Morelia, Michoacán C.P. 58090, México.}%
\author{Omar Gallegos}
 \email{omar.gallegos@iem.cfmac.csic.es}
\affiliation{%
Centro de Ciencias Matemáticas, Universidad Nacional Autónoma de México, UNAM-Campus Morelia, A. Postal 61-3, Morelia, Michoacán C.P. 58090, México.}
\affiliation{%
Instituto de Estructura de la Materia, IEM-CSIC, C/ Serrano 121, 28006 Madrid, Spain}

\date{\today}

\begin{abstract}

The second law of thermodynamics constitutes a fundamental principle of physics, precluding the existence of perpetual motion machines and providing a natural definition of the arrow of time. Its scope extends across virtually all areas of physical theory. Nonetheless, certain systems are known to admit negative absolute temperatures under well-defined conditions, a phenomenon that has been experimentally observed. In this work, we put forward an \textit{extended} version of the first and second laws, which recovers the conventional statement for positive temperatures and extends its applicability to the negative-temperature domain. Illustrative examples from quantum cosmology and Onsager’s vortices are discussed demonstrating the successful applicability of the proposed extension.

\end{abstract}

\maketitle


Thermodynamics is related to every field of physics, from classical to quantum to gravitational systems, which postulates standard laws: $i)$ \textit{The zeroth law} establishes thermal equilibrium between systems. $ii)$\textit{The first law} indicates the conservation law of energy, which can be expressed as $\mathrm{d}E=T\mathrm{d}S-W$, where $E$ is the total energy, $T$ the temperature, $S$ the entropy and $W$ the work done. These can be related to macroscopic (extensive) parameters such as volume $V$ and (intrinsic) external forces such as pressure $P$, $W=P\mathrm{d}V+...$ $iii)$ \textit{The second law} dictates that the change in entropy for an isolated system is always equal to zero or increases, i.e., $\mathrm{d}S\geq 0$, and thus a time arrow is defined. $iv)$ \textit{The third law} prohibits a system from reaching an absolute temperature equal to zero. \newline
Although these laws are (empirically) fundamental in physics and are obeyed by most systems, there exist some systems where the standard second law (SSL) fails; in these examples, a particular effect is observed: they admit negative absolute temperatures (NAT) \cite{Ramsey-NegativeTemp:1956zz,BALDOVIN20211-NegativeTemp,statistical-negative-temp, Onsager-hydrodynamics,Corichi:2025dfe}. We can ask whether for negative absolute temperatures the standard laws of thermodynamics are fulfilled, or whether we need to change or complete them. In particular, this article focuses on the second law.\newline
To delve deeper into this law, we use the definition of temperature, which is related to entropy and goes beyond kinetic energy. In particular, we use the Boltzmann definition, but it is not unique (for example, the Gibbs temperature \cite{Boltzmann-and-Gibbs-entropy})
\begin{align}
    \label{temperature def}
    \dfrac{1}{T}=\left(\dfrac{\partial S}{\partial E}\right)_{X},
\end{align}
where $( \ \ )_X$ indicates that for partial differentiation, the variables in the first law $X$ should be constant. The fact that a Hamiltonian depends on quadratic terms implies that the SSL is valid. However, with this definition for the temperature (\ref{temperature def}), we can obtain NAT systems.\newline
Negative absolute temperature has been observed in systems where the phase space is bounded. Here, a threshold energy $E^*$ can be obtained due to the finite phase space volume and the additional energy allowing a system to explore wider regions of the phase space. We can get three scenarios: $i)$ for $E<E^*$, $ T>0$; $ii)$ for $E=E^*$, $T \rightarrow \pm \infty$; and $iii)$ for $E<E^*$, $T<0$. This behavior is found not only for bounded phase-space systems, but also for systems that obey the discrete non-linear Schrödinger equation \cite{Dicrete-non-linear-SE,DNLSE-2009}.\newline
A counterintuitive statement is the fact that the negative absolute temperature is ``hotter" than the positive absolute temperature. This is because, when a system with $T<0$ and a system with $T>0$ interact, the final result of the thermal equilibrium of the total system goes to a positive temperature. Therefore, the common temperature in nature is positive. To obtain systems with negative absolute temperature, it is necessary to work with perfectly isolated systems with other systems with positive absolute temperature. The change of sign in temperature can be interpreted as a phase transition (See more details in \cite{Ramsey-NegativeTemp:1956zz,BALDOVIN20211-NegativeTemp,statistical-negative-temp,Temperature-in-and-out}).\newline
Here we put forward an {\it extended}  second law (ESL) of thermodynamics that is valid also for NAT systems, which can be read as 
\begin{align}
\label{ASL}
    {\mathrm{sign}({T})}\,\mathrm{d}S\geq 0,
\end{align}
this new form of the second law of thermodynamics includes the SSL when $T>0$, which is the usual case in physics, and extends the regimen of validity when the system admits $T<0$. This ESL can be applied to different cases where the SSL is violated when $T<0$, for instance, the Onsager's vortices \cite{Onsager-hydrodynamics,PhysRevLett.94.054502-vortex,PhysRevA.97.023617-bosevortex}, incompressible fluids \cite{PhysRevLett.113.165302-planarfluid}, nuclear magnetic chains \cite{PhysRev.109.1441-Spintemperature,RevModPhys.69.1-nuclearmag,PhysRevB.97.041301-Spintemperatureverif,PhysRev.81.156-Nuclearspintimes,PhysRev.81.279-nuclear-spin-negativetemp}, lasers \cite{PhysRevLett.85.3365-laser}, cold atoms \cite{PhysRevLett.98.130404-ultracoldatoms,PhysRevLett.81.3108-coldbosonicatoms,PhysRevLett.95.040403-atomicgasesat}, optical waveguides \cite{PhysRev.112.1940}, and quantum cosmology \cite{Corichi:2025dfe}. Some of these examples have experimentally confirmed the negative absolute temperature and $\mathrm{d}S<0$ in the same regime. An interpretation of this sign change for the temperature is as a phase transition. We will apply, as an illustration, the ESL to an effective model in loop quantum cosmology, as well as the Onsager's vortices to show the consistency  within these examples.

\section{Loop Quantum Cosmology}
\label{section:LQC}
Loop Quantum Cosmology (LQC) is the application of techniques in Loop Quantum Gravity (LQG) \cite{Thiemann:2007pyv,Rovelli:2004tv} to a cosmological model. Here, in particular, an isotropic, homogeneous, and flat universe is considered \cite{Ashtekar:2003hd,Ashtekar:2006wn,Ashtekar:2007em}. Loop quantization replaces the classical Big Bang singularity with a quantum bounce. The effective approach from the loop quantum model allows us to compare the semiclassical behavior using continuous variables with the classical description in  standard cosmology \cite{Bojowald:2001xe,Assanioussi:2018hee,Assanioussi:2019iye,Yang:2009fp,Ashtekar:2006wn,Ashtekar:2007em}. The thermodynamics for a flat effective model in LQC can be studied, where the interplay of entropy and temperature is relevant (See more details in \cite{Corichi:2025dfe,Sadjadi:2012wg,Li-Thermo:2008tc,Zhang-Thermo:2021umq}).\newline 
We focus on the generalized second law (GSL), which states that the sum of the change for all contributions to entropy is non negative. The gravitational part $S_g$ makes use of the logarithmic correction to gravitational entropy and the matter part $S_m$ uses the strong energy condition (SEC), namely $\rho+3P\geq 0$ and $\rho\geq 0$ while $\rho +P \geq 0$. Both entropy contributions are in thermal equilibrium.  The change of entropy with respect to cosmic time can be calculated and read as 
\begin{align}
\label{dot Sg}
    \dot{S}_g&=-2\pi \dfrac{\dot{H}}{H^3}(1+\alpha H^2),\\
\label{dot Sm}
    \dot{S}_m&=-\dfrac{4\pi (P+\rho)}{TH^4}\left(H^2+\dot{H}\right),\nonumber\\
    &=4\pi \dfrac{\dot{H}}{H^3}\dfrac{\dot{H}+H^2}{\left( 2H^2+\dot{H}\right)}\left(\dfrac{\rho_c}{\rho_c-2\rho}\right),
\end{align}
where $P$ is the matter pressure, $\rho$ is the energy density, which come from the first law of matter $T\dot{S}_m=V_A\dot{\rho}+(\rho+P)\dot{V}_A$ and $\alpha$ is the parameter due to the logarithmic correction in the gravitational entropy $S_g=A_A/4+\pi\alpha\ln(A_A/4)+\beta$ \cite{Ashtekar:1997yu,Ghosh:2006ph,Ghosh:2004rq,Meissner:2004ju,Engle:2009vc}. Both entropy contributions make use of the apparent area $A_A=4\pi R^2_A$. For a general Friedmann-Lemaître-Robertson-Walker (FLRW) background, the apparent horizon $R_A=1/\sqrt{H^2+\frac{k}{a^2}}$   \cite{Corichi:2025dfe,Hayward:1997jp,Hayward:1998ee,Cai:2005ra,Faraoni:2015ula}. This horizon is background dependent, but it is independent of the theory of gravity with which we work. Thus, the Hubble parameter $H$ is chosen from an effective theory in LQC where the modified Friedmann equation is read as 
\begin{equation}
    \label{eff flat modified Friedmann eq}
    H^2=\dfrac{8\pi}{3}\rho\left(1-\dfrac{\rho}{\rho_c}\right),
\end{equation}
here, $\rho_c$ is the critical density mass, when $\rho=\rho_c$ the quantum bounce occurs, and the quantum effects are negligible when $\rho<<\rho_c$, and the standard Friedmann equation is re-obtained. In addition, $\dot{H}$ changes the sign at $\rho=\rho_c/2$ because we have \cite{Ashtekar:2006wn,Ashtekar:2007em}
\begin{align}
\label{flat acc modif hubble}
    \dot{H}=-4\pi (P+\rho)\left(1-\dfrac{2\rho}{\rho_c}\right).
\end{align}
Furthermore, the temperature $T$ in the gravitational case is associated with the gravity surface $\kappa$
\begin{align}
    \label{negative temperature}
    T=\dfrac{\kappa}{2\pi},
\end{align}
the surface gravity at the apparent horizon of this cosmic model is given by 
\begin{align}
    \label{FLRW surface gravity}
    \kappa&=-\dfrac{R_A}{2}\left(\Dot{H}+2H^2\right),\nonumber\\
    &=-\dfrac{1}{R_A}\left(1-\dfrac{\Dot{R}_A}{2HR_A}\right).
\end{align}
Note that $T>0$ when $\Dot{H}+2H^2<0$, and $T<0$ when $\Dot{H}+2H^2>0$. This is possible since $\dot{H}<0$ for the region $0<\rho<\frac{\rho_c}{2}$ and $\dot{H}>0$ for $\frac{\rho_c}{2}<\rho<\rho_c$. \newline 
The standard analysis of the second law in (quantum) cosmology does not include the possibility of NAT (because $T=\frac{|\kappa|}{2\pi}$). In LQC, the GSL is violated close to the quantum bounce in almost all cases (See more details in \cite{Corichi:2025dfe,Sadjadi:2012wg}). Nevertheless, we can observe that the LQC phase space is effectively bounded. Thus, the threshold energy is located at $\dot{H}+2H^2$, which is when the temperature changes sign. Additionally, the Universe is an example of a perfectly isolated system. Therefore, the effective LQC model can, in principle, admit negative absolute temperatures.\newline 
Every possible value of the logarithmic correction for the gravitational entropy is considered, where the comparison between the \textit{standard} GSL and \textit{an Extended Generalized Second Law} (EGSL) is extensively discussed \cite{Corichi:2025dfe}. However, in this letter, some illustrative results are presented.\newline 
For every possible case, using the SEC, we know that $\dot{H}< 0$ at $0<\rho<\rho_c/2$; then $H^2+\dot{H}>0$ or $H^2+\dot{H}>0$ are plausible. Therefore, the temperature can take both positive and negative absolute values depending on whether $2H^2+\dot{H}< 0$ or $2H^2+\dot{H}> 0$, respectively. Furthermore, $\dot{H}>0$ at $\rho_c/2<\rho<\rho_c$, here $T<0$.   \newline
The case when $\alpha\geq 0$ includes both the case without correction and the case with positive correction \cite{Hod:2004di}, which implies $1+\alpha H^2\geq 0$. When $\dot{H}+H^2>0$, then $\dot{H}+2H^2>0$ and $T<0$, the GSL is valid when $\dot{S}_m\leq\dot{S}_g$. However, under these conditions, the EGSL is fulfilled when $\dot{S}_g\leq\dot{S}_m$. Now, when $\dot{H}+H^2<0$, here, there are two cases: $2H^2+\dot{H}>0$ and $2H^2+\dot{H}<0$. The GSL and the EGSL agree when $T>0$, that is, $\dot{S}_T\geq 0$. Nevertheless, when $2H^2+\dot{H}>0$, the GSL is valid; the EGSL is violated because we have NAT.\newline 
Moreover, close to the quantum bounce at $\rho_c<\rho<\rho_0$, we have $T<0$. Thus, $\mathrm{sgn}(T)\dot{S}_T>0$, namely, the EGSL is valid because of $\dot{S}_T<0$. This is an extension of the GSL when we consider NAT systems. \newline
The validity regions of both GSL and EGSL for every possible value $\alpha$ are discussed in more detail in \cite{Corichi:2025dfe}, also more geometries in effective LQC are studied, not only the flat universe, where the results are similar for different cases and models as it is illustrated in this section. \newline 

The novel result here is the application of the {\it extended generalized second law} (EGSL) in LQC, particularly in the region close to the quantum bounce ($\frac{\rho_c}{2}<\rho<\rho_c$), where the GSL is violated. Since this system admits NAT and the ESL is written as in (\ref{ASL}), the validity domain of the second law of thermodynamics can be extended in LQC not only for the cases where $T>0$ holds, but also for the cases where NAT is present. This sign change of the temperature can be interpreted as a phase transition for the Universe when $\mathrm{sgn}(\dot{H}+2H^2)$ also changes.\newline 
In addition, the gravity surface is related to the attractive or repulsive gravity in a system; in cosmology, there are different phases where this characteristic is important to describe the evolution of the Universe as a whole. A standard interpretation for the region with NAT in a system is that this region is much "hotter" than positive temperatures. This is consistent with the early universe, where the temperature is higher than that of the universe today.\newline 
Finally, although the semiclassical equations in LQC are invariant under time reversal. The second law indicates a time arrow; the only quantity that does not fulfill this symmetry is the change in gravitational entropy over time (\ref{dot Sg}). With this analysis on time reversal symmetry, we can explore the thermodynamic behavior "before" the quantum bounce. In the whole LQC Universe, both branches, there are three transition points. The first two points are because the temperature changes sign "before" and "after" the bounce at $\dot{H}+2H^2$. The temperature definition is invariant under this inversion. Thus, in each branch, the universe has NAT close to the quantum bounce, and the last transition point is located just at the quantum bounce when $H=0$.\newline 
The thermodynamics in cosmology depends on the cosmic horizon, the apparent horizon is used to build this formalism, but if the choose would be different, we find new results.\newline
In what follows, we shall consider another physical system that exhibits non-standard properties regarding the expected thermodynamic behavior.

\section{Onsager's vortices}
\label{Onsager vortices}
The Onsager vortices \cite{Onsager-hydrodynamics} introduced the idea of having NAT systems, where the Euler equation describes the time evolution of a two-dimensional ($x,y$) incompressible ideal flow $\textbf{u}$, which is bounded in a domain with an area $A$
\begin{align}
    \begin{aligned}
& \partial_t \mathbf{u}+(\mathbf{u} \cdot \nabla) \mathbf{u}=-\frac{\nabla P}{\rho}, \\
& \nabla \cdot \mathbf{u}=0,
\end{aligned}
\end{align}
here, $\rho$ and $P$ are the constant density and the constant pressure of the fluid. The vorticity $w$ can be defined in a perpendicular unitary direction $\hat{\textbf{k}}$ to the plane of the flow as follows 
\begin{equation}
    \nabla \times \mathbf{u}=w \hat{\mathbf{k}},
\end{equation}
the vorticity can contain $N$ point-vortices at any time, each one of them with circulation $\Gamma_i$
\begin{equation}
    w(\mathbf{r}, t)=\sum_{i=1}^N \Gamma_i \delta\left(\mathbf{r}-\mathbf{r}_i(t)\right),
\end{equation}
with this, it is possible to build the corresponding Hamiltonian of the system
\begin{equation}
    \mathcal{H}\left(\mathbf{r}_1, \ldots, \mathbf{r}_N\right)=\sum_{i \neq j} \Gamma_i \Gamma_j \mathcal{G}\left(\mathbf{r}_i, \mathbf{r}_j\right)
\end{equation}
where $\mathcal{G}\left(\mathbf{r}_i, \mathbf{r}_j\right)$ is the Green's function. Since the $N$ vortices are in a bounded area $A$, the phase space is enclosed by a constant energy hypersurface $E$, which follows 
\begin{align}
    \Omega(E)=\int_{\mathcal{H}<E} d q_1 \cdots d q_N d p_1 \cdots d p_N \leq C_N A^N,
\end{align}
where $C_N=\prod_{i=1}^N\left|\Gamma_i\right|$. In addition, the degrees of freedom are constituted by the canonical coordinates of the $N$ point vertices, they are given by $q_i=\sqrt{\left|\Gamma_i\right|} x_i, \quad p_i=\operatorname{sign}\left(\Gamma_i\right) \sqrt{\left|\Gamma_i\right|} y_i$.
Furthermore, entropy of this system is calculated using the density of states $\omega(E)=\mathrm{d}\Omega(E) / \mathrm{d} E$. Thus,
\begin{equation}
    S(E)=k_B \ln \omega(E),
\end{equation}
where $k_B$ is the Boltzmann constant. In addition, the density of states approaches zero when $E\rightarrow \pm \infty$. There exists a maximum finite value $E_M$, where the entropy decreases to obtain NAT at $E>E_M$. In this case, the application of ESL at $E>E_M$, when we have NAT, extends the validity domain because the SSL is violated, but ESL is valid.\newline
This line of investigation has been addressed in many articles using analytical \cite{Onsager-hydrodynamics,BALDOVIN20211-NegativeTemp,statistical-negative-temp}, numerical \cite{PhysRevLett.94.054502-vortex,PhysRevLett.113.165302-planarfluid,PhysRevA.97.023617-bosevortex}, and experimental \cite{expNAT1,expNAT2} methods, in which the NAT and SSL violations have been confirmed.
\section{Conclusions}
\label{section:Conclusion}

The second law of thermodynamics is generally valid for systems at positive absolute temperature. However, it is natural to ask whether this standard second law remains applicable in situations where a system admits negative absolute temperatures. Such systems, studied across various areas of physics, display unconventional thermodynamic behavior and, in some cases, have been observed to violate the SSL.

We have reviewed the conditions under which NAT can occur, following earlier foundational work~\cite{Ramsey-NegativeTemp:1956zz,BALDOVIN20211-NegativeTemp,statistical-negative-temp}. These conditions include the assignment of a temperature to each thermodynamic state, the existence of a bounded phase space, and the isolation of the system from others at positive temperature.

In this manuscript, we have proposed an extension of the second law of thermodynamics, which reproduces the standard second law for $T>0$ and remains valid for systems with $T<0$. This extended formulation is expressed in Eq.~(\ref{ASL}). We have successfully illustrated its applicability with two representative examples exhibiting negative absolute temperature, showing that the extended second law enlarges the domain of validity compared to the SSL. In both cases, the systems satisfy the NAT conditions: they are isolated and possess a bounded phase space that leads to the existence of a threshold energy.

The systems discussed in this article originate from distinct physical fields, suggesting that this extension of the second law may possess broad, potentially universal applicability. Future research could test the ESL across a wider class of systems exhibiting NAT, for which the SSL fails to account for the observed "paradoxical" thermodynamic behavior \cite{Onsager-hydrodynamics,PhysRevLett.94.054502-vortex,PhysRevA.97.023617-bosevortex,PhysRevLett.113.165302-planarfluid,PhysRev.109.1441-Spintemperature,RevModPhys.69.1-nuclearmag,PhysRevB.97.041301-Spintemperatureverif,PhysRev.81.156-Nuclearspintimes,PhysRev.81.279-nuclear-spin-negativetemp,PhysRevLett.85.3365-laser,PhysRevLett.98.130404-ultracoldatoms,PhysRevLett.81.3108-coldbosonicatoms,PhysRevLett.95.040403-atomicgasesat,PhysRev.112.1940,Corichi:2025dfe}.

While the interpretation of negative temperature remains unsettled, we expect that this simple formulation will contribute to a clearer understanding of its thermodynamic role and stimulate further research.

\section{Acknowledgments}
\label{section:Acknowledgments}
We thank Prof. Hugo A. Morales-Técotl for the discussions and suggestions. O.G. thanks that this work was supported by UNAM Posdoctoral Program (POSDOC). Additionally, the work of O.G. was financially supported by SECIHTI postdoctoral fellowships and SNII with CVU 786532.

\bibliography{ref}

\end{document}